\documentclass[aps,prl,twocolumn,superscriptaddress]{revtex4-2}
\usepackage{graphicx}
\usepackage{graphics}
\usepackage{amsmath}
\usepackage{amssymb}
\usepackage{amsfonts}
\usepackage{dsfont}
\usepackage{braket}
\usepackage{color}
\usepackage{braket,slashed}
\usepackage[mathscr]{euscript}
\definecolor{darkblue}{rgb}{0, 0, 0.8}
\usepackage[colorlinks=true, breaklinks=true, linkcolor=red, citecolor=blue, urlcolor=blue]{hyperref} 
\usepackage{hyperref}
\usepackage{subfigure}
\usepackage{xfrac}
\usepackage{bm}
\usepackage{kantlipsum}
\usepackage{enumitem}
\usepackage{tikz}
\usepackage{framed}
\usepackage{graphicx}
\usepackage{subfigure}
\usepackage{cleveref}
\usepackage{array}

\newcommand{\parL}[1]{\noindent\textbf{\textit{#1}}---}

\allowdisplaybreaks[1]

\newcommand{\code}[1]{\texttt{#1}}

\newcommand{\e}{\ensuremath{\mathrm{e}}}

\newcommand{\textmath}[1]{\ensuremath{\text{\textit{#1}}}}

\newcommand{\Z}{\ensuremath{\mathbb{Z}}}
\newcommand{\U}{\ensuremath{\mathrm{U}}}

\begin{document}

\title{Interface roughening in the 3-D Ising model with tensor networks}

\author{Atsushi Ueda}
\affiliation{Department of Physics and Astronomy, Ghent University, Krijgslaan 281, 9000 Gent, Belgium}

\author{Lander Burgelman}
\affiliation{Department of Physics and Astronomy, Ghent University, Krijgslaan 281, 9000 Gent, Belgium}

\author{Luca Tagliacozzo}
\affiliation{Institute of Fundamental Physics IFF-CSIC, Calle Serrano 113b, Madrid 28006, Spain}

\author{Laurens Vanderstraeten}
\affiliation{Center for Nonlinear Phenomena and Complex Systems, Universit\'e Libre de Bruxelles, CP 231, Campus Plaine, 1050 Brussels, Belgium}

\date{\today}

\begin{abstract}
Interfaces in three-dimensional many-body systems can exhibit rich phenomena beyond the corresponding bulk properties. In particular, they can fluctuate and give rise to massless low energy degrees of freedom even in the presence of a gapped bulk. In this work, we present the first tensor-network study of the paradigmatic interface roughening transition of the 3-D Ising model using highly asymmetric lattices that are infinite in the $(xy)$ direction and finite in $z$. By reducing the problem to an effective 2-D tensor network, we study how truncating the $z$ direction reshapes the physics of the interface. For a truncation based on open boundary conditions, we demonstrate that varying the interface width gives rise to either a $\mathbb{Z}_2$ symmetry breaking transition (for odd $L_z$) or a smooth crossover(for even $L_z$). For antiperiodic boundary conditions, we obtain an effective $\mathbb{Z}_q$ clock model description with $q=2L_z$ that exhibits an intermediate Luttinger liquid phase with an emergent $\U(1)$ symmetry.
\end{abstract}

\maketitle

\parL{Introduction}%
%
Tensor networks have been proven to be a powerful tool to study the phase diagrams of strongly-correlated models in statistical mechanics. They have emerged as a complementary approach to traditional computational methods such as Monte Carlo sampling. A key advantage is that tensor networks can access bulk properties directly in the thermodynamic limit, circumventing the need for finite-size extrapolations.
\par Yet, many-body systems often exhibit interesting phenomena that cannot be captured by the bulk properties alone. One example is the physics of interfaces, which can undergo abrupt changes in their qualitative behavior independently from the properties of the bulk of the system. A classic example is the so-called rough phase and the corresponding roughening transition appearing in 3-D systems, which represent the softening of the transverse mode of the interface, and as such at low energies is described by a 2-D Berezinskii-Kosterlitz-Thouless (BKT) universality class \cite{Berezinskii1971, Berezinskii1972, Kosterlitz1973, Kosterlitz1974}. Numerically, such transitions are notoriously challenging due to the logarithmic finite-size corrections. When combined with the difficulty for simulating 3-D systems, this leads to particularly formidable numerical challenges. 
\par In this work, we demonstrate how tensor networks can be used to study interface transitions and phases in 3-D statistical mechanics. The main idea is that such transitions and phases are already visible for a very small size of the system transverse to the interface. We can thus consider an infinite slab of finite width, or wrap the finite width on a small circle. After grouping a few tensors, we are back to the standard infinite 2D lattices which we can treat with the usual tensor network techniques. In order to showcase these ideas, we will focus on the paradigmatic example of the roughening transition in the ferromagnetic Ising model on the simple cubic lattice. In this way we can characterize the effects of the compact third dimension on the well-known physics of interfaces. In particular we will investigate how the transition to the rough phase is affected by this truncation from both numerical and theoretical point of views.

\parL{Interface in the 3-D Ising model}%
%
We study a classical Ising model on a cubic lattice. The energy of each spin configuration $\{s_i\}$ is given by the ferromagnetic Ising Hamiltonian
\begin{equation} \label{eq:H}
    \mathcal{H}_{\textmath{Ising}}(\{s_i\}) = - J \sum_{\braket{ij}} s_i s_j, \quad s_i \in \{+1,-1\},
\end{equation}
where the sum runs over all nearest-neighbor bonds. This model undergoes a bulk phase transition from a disordered phase at high temperature to an ordered phase at low temperature. The best estimates for the corresponding critical temperature are around $T^c_{\textmath{bulk}}\approx4.512$ \cite{Talapov1996}. 

Once one of the direction is taken finite and compact, the 3D Ising transition turns into a 2D Ising transition, which is interpreted as a finite temperature transition of the corresponding 2D quantum Ising model in a transverse field. The actual size at which this transition takes place varies depending on the $\beta$ we use to define the partition function $Z=\sum_{\{s\}} \exp(-\beta \mathcal{H})$.  

The low-temperature phase spontaneously breaks the $\Z_2$ symmetry of the Ising model, giving rise to long-range order in the system with all spins pointing either predominantly up or down. This leaves the possibility of realizing interface configurations that interpolate between both bulk orders \cite{Dobrushin1973, Weeks1973}. One possibility is to take a an infinite slab with finite thickness. For example we fix all the spins on the upper face of the slab up, and on the lower face down. Somewhere in the middle of the slab, a surface of spin up will be adjacent to a surface of spin down, creating an interface. At sufficiently low temperatures, the interface is well-localized in the $z$-direction due to the ferromagnetic interactions in space and it behaves as a smooth surface. By increasing the temperature, however, it can fluctuate in space, and when fluctuations are sufficiently large, it looses its 2-D geometry and becomes more like a fractal object. This is the temperature which identifies the roughening transition into what is called a rough phase. When the system is infinite in all directions, this roughening transition occurs at a temperature around $T^c_{\textmath{int}}\approx2.454$ \cite{Mon1988, Mon1990, Hasenbusch1996, Hasenbusch1997}, well below the bulk critical temperature.

The physics of this roughening transition can be described effectively by a 2-D solid-on-solid (SOS) model \cite{Chui1976, VanBeijeren1977, Chui1978, Weeks1980}. The idea is that we start from a reference interface at a certain location in space, and we describe the fluctuations of the interface by height variables $h_i\in\Z$ on a square lattice. The effective Hamiltonian is given by 
\begin{equation} \label{eq:sos}
    \mathcal{H}_{\textmath{SOS}}(\{h_i\}) = \sum_{\braket{ij}} V(h_i-h_j),
\end{equation}
describing the energy cost of having different interface locations on neighboring sites. The fact that each term only depends on the difference between the heights, reflects that the location of the reference interface is arbitrary. As a result, the SOS model exhibits a microscopic $\U(1)$ symmetry. While the most natural choice for the energy functional is $V(h)=|h|$, a variety of SOS models with the same universal properties can be retrieved by different choices for $V(h)$. For the particular choice $V(h)=h^2$, the SOS model is dual to the Villain model \cite{Villain1975}, which itself is a variation of the classical XY model \cite{Jose1977}. The latter is known to exhibit a BKT-type phase transition \cite{Berezinskii1971, Berezinskii1972, Kosterlitz1973, Kosterlitz1974}, suggesting that the roughening transition is also of that type \cite{Chui1976}. The BKT nature of the roughening transition was confirmed by large-scale Monte-Carlo simulations \cite{Mon1988, Mon1990, Hasenbusch1996, Hasenbusch1997}.

\begin{figure}
    \centering
    \includegraphics[width=0.99\columnwidth,page=2]{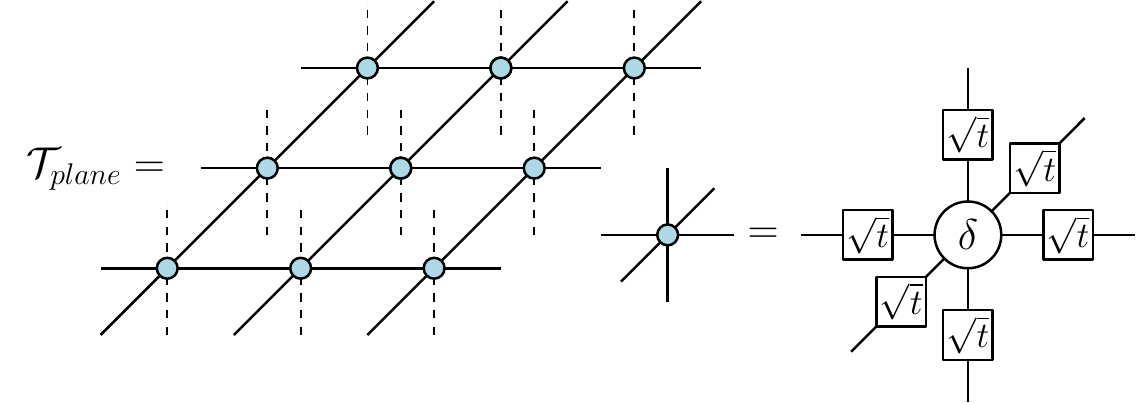}
    \caption{\textit{Plane-to-plane transfer matrix.} The tensor is defined in terms of the six-legged $\delta$ tensor [Eq.~\eqref{eq:delta}] and the square root of the $t$ matrix [Eq.~\eqref{eq:t_matrix}] encoding the elementary Boltzmann weights.}
    \label{fig:T_plane}
\end{figure}

\parL{Tensor network approach}%
%
The Ising partition function can be represented as an infinite 3-D tensor network in terms of the six-legged tensors
\begin{equation} \label{eq:delta}
    \delta_{ijklmn} = \left\{ \begin{array}{l} 1 \quad \textmath{if}\;i=j=k=l=m=n \\
    0 \quad \textmath{else} \end{array} \right.
\end{equation}
on the sites and the $t$ matrices on the links representing the local Boltzmann weights
\begin{equation} \label{eq:t_matrix}
    t = \begin{pmatrix} \e^{+\beta} & \e^{-\beta} \\ \e^{-\beta} & \e^{+\beta} \end{pmatrix} \;.
\end{equation}
This infinite 3-D contraction can be broken down as the infinite power of the plane-to-plane transfer operator $\mathcal{T}_\textmath{plane}$, which is a tensor-network operator acting on an infinite square lattice of two-level systems (see Fig.~\ref{fig:T_plane}). The contraction problem can be formulated as a fixed point equation for this transfer operator. This fixed point can itself be approximated as an infinite projected entangled-pair state (PEPS), which can be optimized as a variational optimization problem for the PEPS tensors \cite{Nishino2000, Vanderstraeten2018, Vanhecke2022, Xu2025b}. As discussed already in several places, the advantage of using tensor networks with respect to standard Monte Carlo methods in this context, is they provide a direct access to the free energy of the system, something that in Monte-Carlo requires sophisticated out-of-equilibrium techniques based on the Jarzynski inequalities \cite{jarzynski1997}. In the tensor network setting, the cost function that is optimized is the free energy density, so the PEPS approach provides an immediate variational estimate for the free energy. Once the PEPS fixed point is found, all other local expectation values and correlation functions can be computed as well.

\begin{figure}
    \centering
    \includegraphics[width=0.8\columnwidth, page=1]{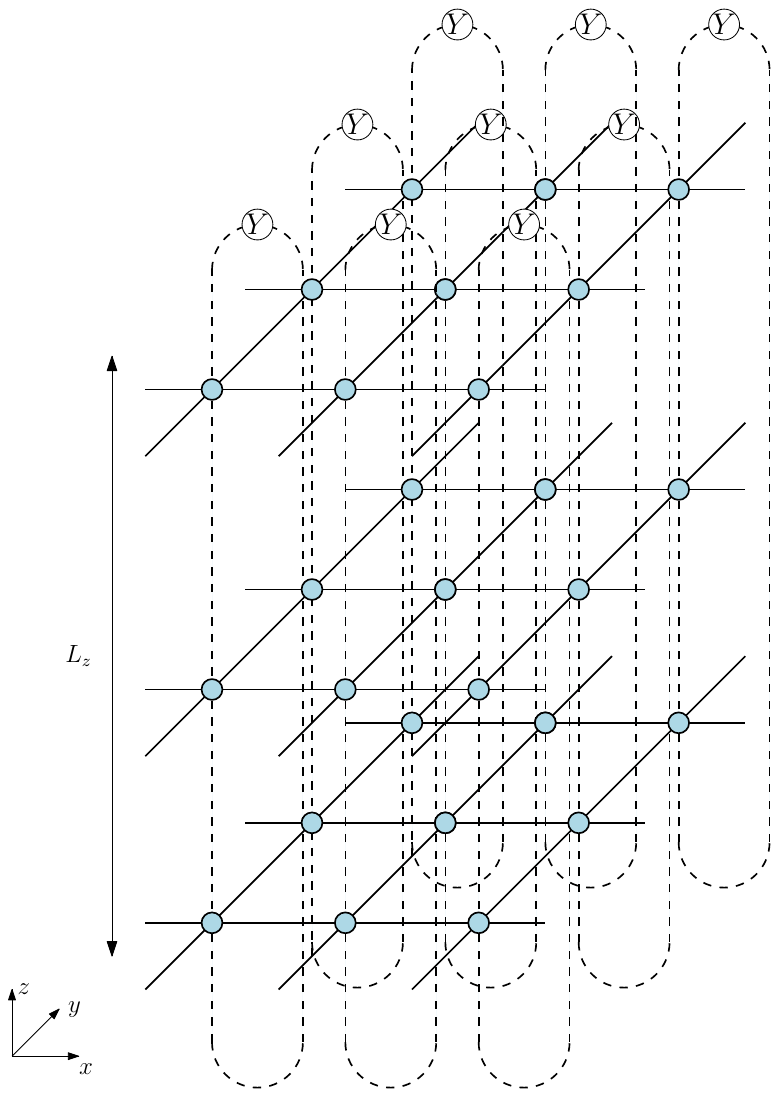}
    \caption{\textit{The interface partition function.} We start from $L_z$ layers of the plane-to-plane transfer matrix, and trace with matrices $Y$ that define the boundary conditions. }
    \label{fig:interface}
\end{figure}

For studying the properties of an interface we work with finite geometries.  The system is infinite in two directions, the $(xy)$ plane, and finite along the $z$ direction with an extent $L_z$. The partition function on this slab geometry is then obtained as the product of a finite number of $\mathcal{T}_\textmath{plane}$. We can either wrap the third direction periodically, or add any desired twist to the boundary conditions in the $z$ direction. This is done by inserting in the tensor network  a layer of matrices $Y$ before taking the trace. Different choices of $Y$ matrices implement different boundary conditions, see Fig.~\ref{fig:interface}. The standard periodic boundary conditions corresponds to choosing all $Y$ as identity matrices. An interface is created by choosing the $Y$ that creates a cut in the $z$ direction and fixes the spins to be up/down on the top/bottom of the slab thus realizing fixed open boundary conditions of opposite nature. The corresponding $Y$ is given in Eq.~\eqref{eq:Y_obc}. Alternatively we can build antiperiodic boundary conditions choosing $Y$ as in Eq.~\eqref{eq:Y_tbc}, which flips the sign of the interactions on these links. In all cases, the $z$-direction of the tensor network can be compactified such that the partition function corresponds to the contraction of a 2-D tensor network with a virtual dimension $D=2^{L_z}$. We use two different techniques for contracting this infinite 2-D tensor network and extracting its critical properties.

On the one hand, we use variational boundary MPS methods to find the fixed point of its row-to-row transfer matrix \cite{Haegeman2017, Fishman2017} using the VUMPS algorithm \cite{ZaunerStauber2018, Vanderstraeten2019}. Here, the control parameter is the bond dimension $\chi$ of the boundary MPS. We can calculate the magnetization profile of the interface, as well as the correlation length $\xi$ in the $(xy)$ plane. Moreover, in a critical phase we can extract the central charge $c$ by fitting the correlation length $\xi_\chi$ and entanglement entropy $S_\chi$ of the boundary MPS with different bond dimension $\chi$, according to the scaling formula \cite{Tagliacozzo2008, Pollmann2009} $S_\chi \propto \frac{c}{6} \log \xi_\chi $.

\begin{figure}
    \centering
    \includegraphics[width=\linewidth]{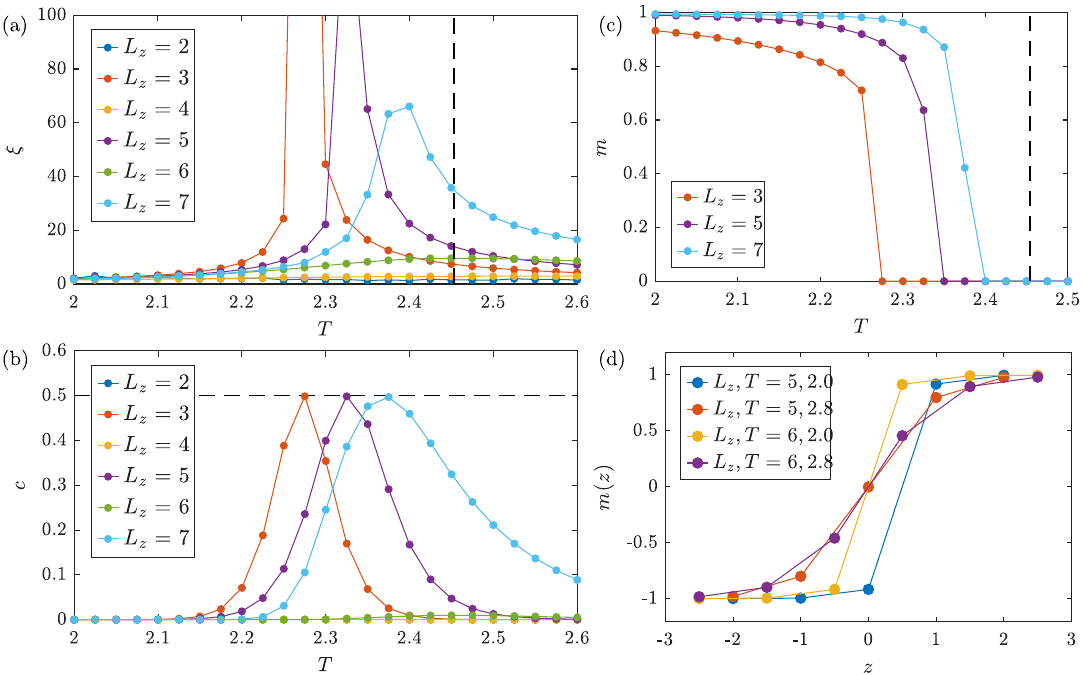}
    \caption{\textit{Numerical results for open boundary conditions.} (a) Boundary MPS correlation length, (b) central charge from TNR simulations, and  (c)total magnetization as a function of temperature for different values of $L_z$ (for even $L_z$ the magnetization is always zero within machine precision);(d) magnetization profiles for a well chosen points from boundary MPS. In all these simulations the boundary MPS bond dimension is $\chi=64$. The dashed line is the Monte Carlo estimate for the roughening transition temperature \cite{Hasenbusch1997}.}
    \label{fig:obc}
\end{figure}

On the other hand, we make use of a tensor network renormalization (TNR) approach \cite{Evenbly2015, Evenbly2017, Gu2009, Hauru2018, Yang2017, Bal2017, Homma2024, Ueda2025}. In this approach, the two-dimensional network is contracted via a sequence of coarse-graining steps on the rank-4 local constituent tensor. The coarse-grained tensor after $n$ renormalization steps provides a direct representation of a subsystem of linear size $L = 2^{n/2}$ in the $(xy)$ plane. Using the classical-quantum correspondence, the logarithm of the transfer matrix spectrum made by the renormalized tensor can be considered as finite-size energy spectrum. Their analysis as a function of $L$ then gives direct access to the critical properties of the system, such as the central charge and the scaling dimensions of primary operators and running couplings~\cite{Gu2009, Ueda2023}. In particular, the thermal TNR approach~\cite{Ueda2025} combines the conventional TNR with the compression of bonds in $z$-direction~\cite{Czarnik_2015,Czarnik_2016}, which allows us to simulate systems with larger number of layers. 
\parL{Open boundary conditions}%
%
We start by investigating the partition function with open boundary conditions, obtained by the tensor network in Fig.~\ref{fig:interface} with the matrix
\begin{equation} \label{eq:Y_obc}
Y_{\textmath{obc}} = \begin{pmatrix} 0 & 1 \\ 0 & 0 \end{pmatrix}
\end{equation}
on the traced bonds. In this setup the partition function is invariant under a combination of (i) a spatial reflection in the $z$ direction and (ii) the flipping of all spins. The reflection is bond-centered for even $L_z$ and site-centered for odd $L_z$. In Fig.~\ref{fig:obc} we have plotted (a) the correlation length, (b) the effective central charge and (c) the total magnetization per column as a function of $T$ and $L_z$, as well as (d) the interface magnetization profile for a few well-chosen points. These results show a qualitative difference between even and odd values of $L_z$.

In the case of even $L_z$, we find that the correlation length increases as we approach the roughening transition and reaches a plateau at higher temperatures; the height of the plateau increases as we increase the number of layers. Similarly, in the magnetization profile we see that the interface gets more extended as the temperature crosses the roughening transition. This behavior is not unexpected: We can interpret the effect of the open boundaries as pinning the interface to sit in the middle of the system. This breaks the $\U(1)$ symmetry of the SOS model and gaps out the entire BKT phase, leading to a crossover between smooth and rough phases.

For an odd number of sites, however, the choice of the location of the interface is ambiguous and breaks the site-centered reflection symmetry of the model. Inspecting the magnetization profiles numerically, we indeed find that the interface location is chosen spontaneously at low temperature, leading to a non-zero total magnetization in the system. For larger temperatures, however, we find that the symmetry is restored by setting the magnetization to zero in the middle of the slab. We find, therefore, a continuous, $\Z_2$-symmetry breaking, phase transition with diverging correlation length between the two phases.

These considerations are confirmed numerically as shown in Fig.~\ref{fig:obc}, where we have plotted the scaling properties from boundary MPS (d) and TNR simulations(e). We find clear evidence for a $\Z_2$ symmetry breaking phase transition with central charge $c=1/2$ only for odd $L_z$, and no trace of a critical scaling in the high-temperature regime. We expect to find the critical signatures of the rough phase only for higher values of $L_z$.

\begin{figure}
    \centering
    \includegraphics[width=\columnwidth]{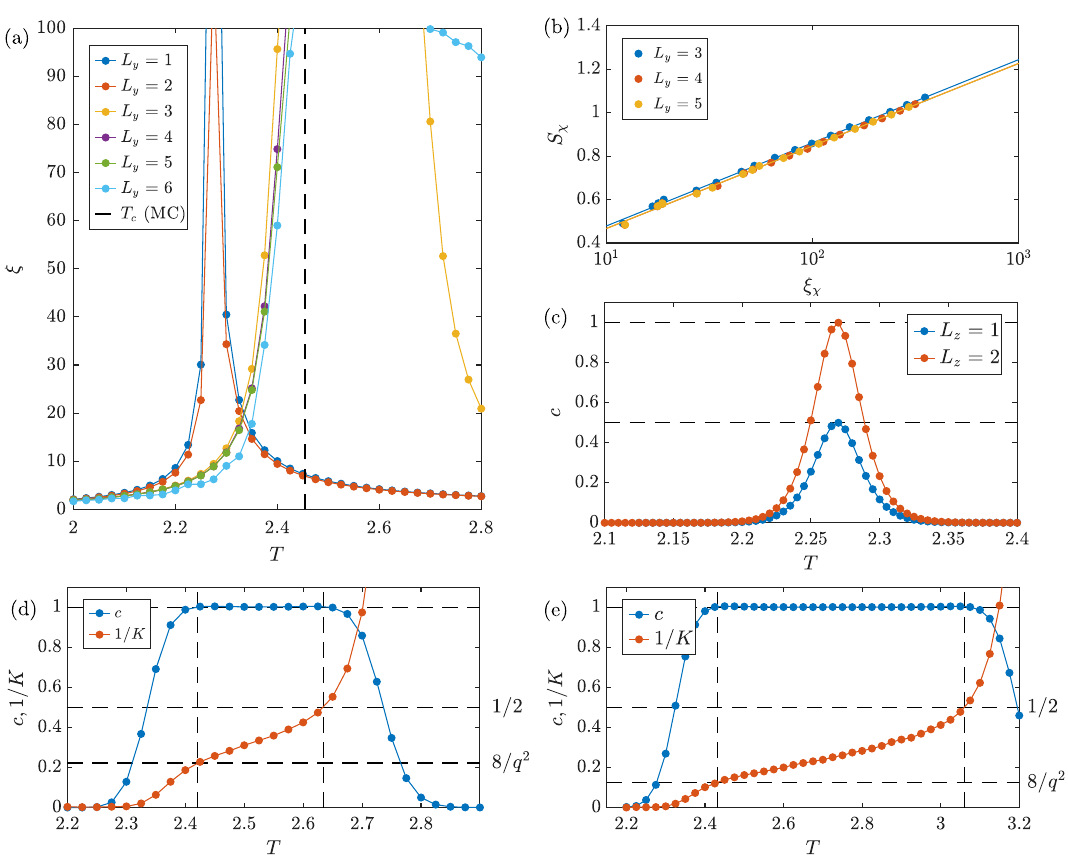}
    \caption{\textit{Numerical results for antiperiodic boundary conditions.} (a) Correlation length from boundary MPS calculations at $\chi=64$, (b) the scaling of the MPS entanglement entropy as a function of the correlation length at $T=2.5$ for three different values of $L_z$, fitted with the form $S_\chi\propto c/6\log\xi_\chi$ (for all three fits we find $c=1$ within one percent error). The temperature dependence of the central charge and Luttinger parameter computed from TNR for (c), $ L_z=1$(cross) and 2(diamond), (d)$, L_z=3$ and (e)$\,L_z=4.$ The vertical dashed lines correspond to $T_c^{(1)}$ and $T_c^{(2)}$. $T_c^{(1)}$ corresponds to the roughening transition. In the TLL regime concerned, the lowest excitation corresponds to $W_1=:e^{i\theta}:$, whose scaling dimension is $\Delta_{W_1}=1/4K$. We therefore extract the Luttinger parameter $K$ from $\Delta_{W_1}$. The Luttinger parameters (red circles) take the universal values at the BKT transition points.}
    \label{fig:tbc}
\end{figure}

\parL{Antiperiodic boundary conditions}%
%
Next, we examine the case of antiperiodic boundary conditions, implemented by choosing the $Y$ matrix
\begin{equation} \label{eq:Y_tbc}
Y_{\textmath{abc}} = \begin{pmatrix} 0 & 1 \\ 1 & 0 \end{pmatrix}
\end{equation}
on the traced bonds in Fig.~\ref{fig:interface}. In this geometry, frustration arises along the $z$-direction: While the ferromagnetic interaction favors aligned spins, there is always a bond that creates a an interface due to the antiperiodic boundary condition. In Fig.~\ref{fig:tbc} we show the numerical results. For $L_z=1,2$ we again find again a single $\Z_2$ symmetry breaking phase transition, but for larger $L_z$ we find a critical phase that was absent in the case of open boundary condition.

\begin{figure}
    \centering
    \includegraphics[width=0.85\columnwidth, page=3]{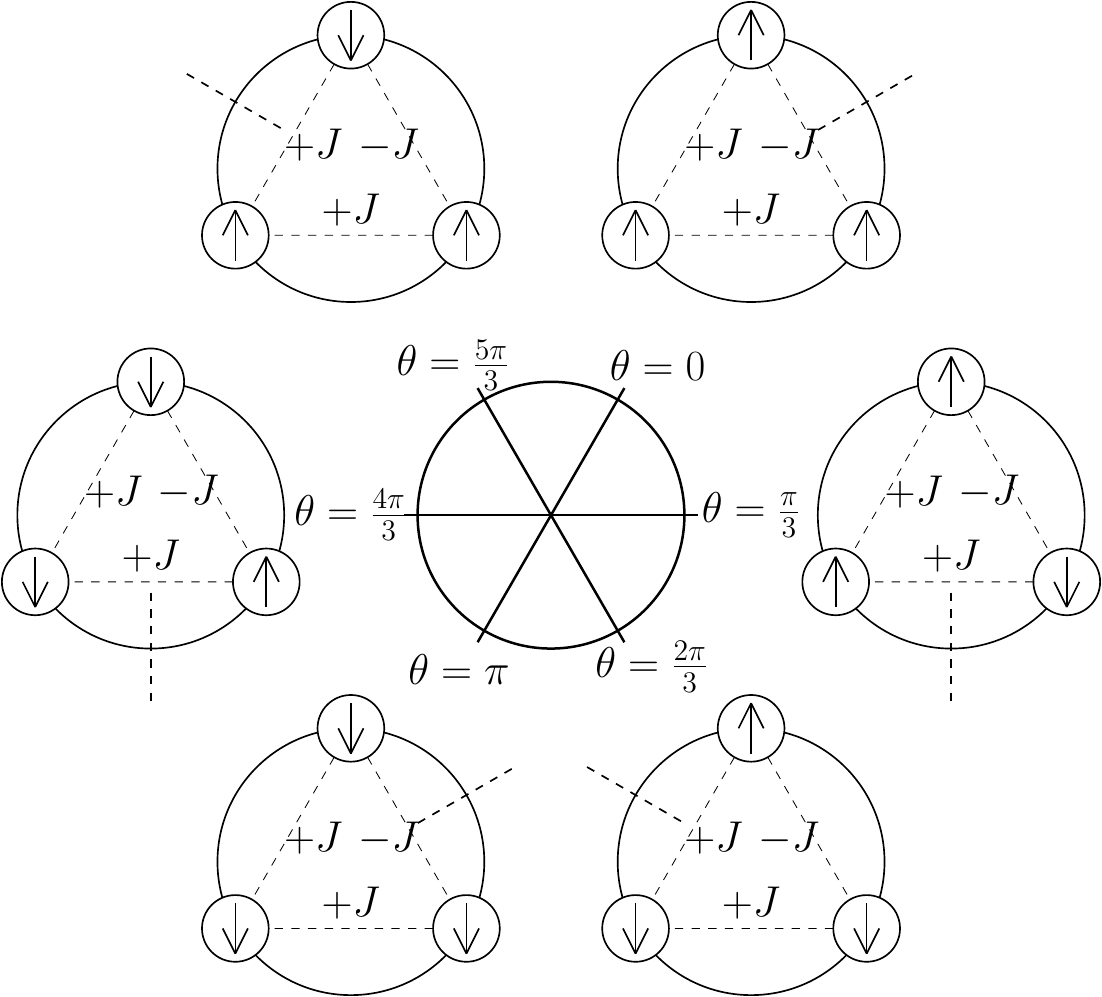}
    \caption{\textit{Clock model interpretation.} The six local interface configurations on a single column of the 3-D Ising model with $L_z=3$ and antiperiodic boundary conditions. The antiperiodic bond is denoted by antiferromagnetic interaction $(-J)$, the other two bonds have a ferromagnetic interaction $(+J)$. The dashed line denotes the location of the interface: either two parallel spins on the antiperiodic bond, or two antiparallel spins on a normal bond. In the middle, we show the identification of these six configurations with a clock variable $\theta=2\pi k /(2L_z)$.}
    \label{fig:clock}
\end{figure}

In order to explain this behavior, we will write down an effective SOS-like model, similar to Eq.~\eqref{eq:sos} above. First we note that the local interface configurations on a single column can be sequentially generated by moving it clockwise through the $z$ direction, as illustrated for a three-site system in Fig.~\ref{fig:clock}. Consequently, we obtain $q=2L_z$ configurations, which we label using a discrete angle $\theta=2\pi k/q$ with $k\in\{0,1,\cdots,q-1\}$. In analogy to Eq.~\eqref{eq:sos}, we define a model on the square lattice with on each site $i$ of the $(xy)$ plane a set of domain-wall configurations $\{\theta_i\}$, and with a Hamiltonian of the form
\begin{equation}
    \mathcal{H}_{\textmath{eff,tbc}} = \sum_{\langle ij\rangle}W(\theta_i,\theta_j).
\end{equation}
The interaction between two adjacent sites from the planar Ising interaction is $-L_z$ if $\theta_i=\theta_j$ coincide. More generically, the interaction should be of the form
\begin{align}
    W(\theta_i,\theta_j) = -\frac{q}{2} + \left|\frac{q}{\pi}[\theta_i-\theta_j]_{2\pi}\right|,\label{eq:interaction}
\end{align}
where $[\theta_i]_{2\pi}$ is a $2\pi$ modulo function that takes a value in $(-\pi,\pi]$. Note that this interaction exhibits a microscopic $\mathbb{Z}_{q}$ symmetry: The interaction only depends on the difference in $\theta$s, and is thus invariant under rotations. Moreover, this $\mathbb{Z}_{q}$ symmetry is present not only in the low-energy interface subspace, but in \emph{all} configurations. This can be seen by moving the $Y$ matrices [Eq.~\eqref{eq:Y_tbc}], which are symmetries of the tensors, through the network without changing the value of the partition function.

Exploiting this, we can expand the effective nearest-neighbor interaction in [Eq.~\eqref{eq:interaction}] in the $\mathbb{Z}_q$-symmetric basis as 
\begin{equation}
W(\theta_i,\theta_j) =
-\sum_{\substack{k=1\\ k:\mathrm{odd}}}^{q/2-1}
\frac{4}{q\,\sin^2\!\Bigl(\frac{\pi k}{q}\Bigr)}
\cos\Bigl(k(\theta_i-\theta_j)\Bigr) \;.
\end{equation}
This has the form a generalized clock model, for which the phase diagram and critical behaviour are rather well-known \cite{Jose1977, Wiegmann1978, Elitzur1979, Cardy1980, Alcaraz1980, Ortiz2012}. At small temperature, the system selects one of the interface configurations, and the $\mathbb{Z}_q$ symmetry is spontaneously broken. For $q\leq4$ the system restores the symmetry at higher temperature through a single critical point with central charges $c=1/2,1$ for $q=2,4$ respectively. For larger $q>4$, as the temperature increases a spin-wave description
\begin{equation} \label{eq:ham_clock}
    \mathcal{H}_{\textmath{eff,tbc}} \sim - \frac{\ln 2}{\pi^2}q^2\sum_{\langle ij\rangle}(\theta_i-\theta_j)^2.
\end{equation}
becomes valid at $T_c^{(1)}$. Finally, there is another transition at $T_c^{(2)}$ into a symmetric regime at high-temperatures, happening when the vortices, formed by the rotors, begin to proliferate due to their entropy gain.

The critical phase exhibits an emergent $\U(1)$ symmetry and is described by a Tomonaga-Luttinger liquid field theory \cite{Kadanoff1979, Luttinger1963, Tomonaga1950} with Hamiltonian
\begin{equation}
    H_{\mathrm{TLL}} = \int dx\left[\frac{K}{2\pi}(\partial_x\theta)^2 +\frac{1}{2\pi K}(\partial_x\phi)^2\right],\label{eq:tll}
\end{equation}
where $\phi$ represents a dual field of $\theta$ compactified as $\phi\sim\phi+\pi$, and where $K$ is the Luttinger parameter. The transition into the low-temperature phase at $T_c^{(1)}$ is then understood to occur when the $\Z_q$ symmetry-breaking field in the continuum theory becomes relevant, characterized by $K=q^2/8$. The transition into the high-temperature phase at $T_c^{(2)}$ occurs when the vortex-pair creation operator, $V_1=:\cos(2\phi):$, becomes relevant; this transition is characterized by $K=2$. 

A rough estimate for the critical temperatures can be obtained by inserting [Eq.~\eqref{eq:ham_clock}] into [Eq.~\eqref{eq:tll}], giving $T_c^{(1)}\sim 16\ln(2)/\pi\simeq 3.53$ and $T_c^{(2)}\propto q^2$. Although the interface expansion (Fig.~\ref{fig:clock}) is no longer quantitatively reliable in this temperature regime, these simple analytical estimate are consistent with the observation that, as $q\to\infty$, the first transition point $T_c^{(1)}$ becomes $q$-independent, whereas the second transition is pushed to ever increasing temperatures (interrupted by the bulk Ising transition in $L_z\rightarrow\infty$). The first transition point can therefore be identified with the roughening transition as $L_z\to\infty$, whereas the second transition is an artifact of finite $L_z$.

To test this low-energy picture, we simulate the truncated interface model of Fig.~\ref{fig:interface} with boundary condition Eq.~\eqref{eq:Y_tbc} using boundary MPS and TNR.  The boundary MPS calculations show that for $L_z=1,2$, the system undergoes a single transition, while for $L_z\geq3$ an extended critical temperature window opens up. The critical theory is confirmed to be $c=1$ CFT as shown in (a-b).

The TNR data provide a sharper fingerprint. From the transfer-matrix spectrum of the renormalized tensors, we extract the central charge and the scaling dimensions of low-lying operators. Using that the lowest-scaling dimension corresponds to $\Delta_{W_1} = 1/(4K)$ of $W_1 = :e^{i\theta}:$, we determined the Luttinger parameter $K$. As shown in (c-e), $K$ approaches the universal values at the two BKT transitions $T_c^{(1)}$ and $T^{(2)}_c$ estimated from the central charge. This corroborates our generalized clock-model interpretation.

\begin{figure}
    \centering
    \includegraphics[width=\columnwidth]{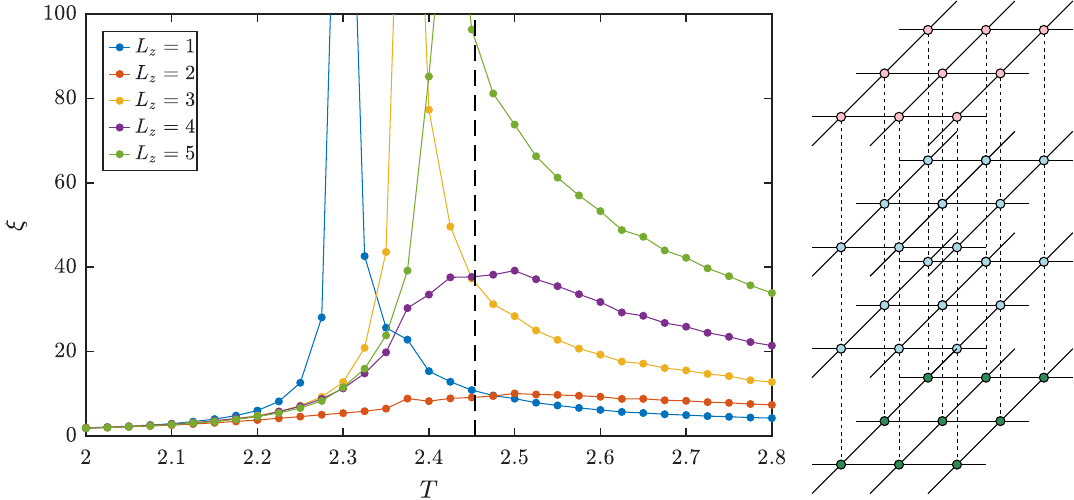}
    \caption{\textit{Numerical results for PEPS boundary conditions.} The correlation length as a function of temperature for different values of $L_z$, where we have inserted $L_z$ layers of the plane-to-plane transfer operator in between the two symmetry-broken PEPS fixed points.}
    \label{fig:peps}
\end{figure}

\parL{PEPS boundary conditions}%
%
Finally, we can also use the PEPS fixed points of the plane-to-plane transfer operator of Fig.~\ref{fig:T_plane} as boundary conditions. Using variational optimization \cite{Vanderstraeten2018}, we have found PEPS fixed points with bond dimension $D=2$ (which is highly accurate in this temperature regime), inserted a number of layers $L_z$ in between the two distinct symmetry-broken PEPS fixed points, and contracted the resulting 2-D tensor network (see the right panel of Fig.~\ref{fig:peps}). We find essentially the same behavior as in the case of open boundary conditions, but the critical points for odd $L_z$ are shifted closer to the 3-D value for the roughening transition temperature and the correlation lengths are larger in the high-temperature regime. We find that the PEPS fixed points provide smoother boundaries than taking simple open boundary conditions. In cases where the 3-D bulk correlation length is larger, the use of these PEPS boundary conditions will clearly yield a more precise approximation of the interface partition function.

\parL{Outlook}%
%
In this work, we have shown that interfaces in 3-D statistical mechanics can be simulated efficiently with tensor networks. To this end, we compactify the $z$ direction to a slab of thickness $L_z$ and then contract the resulting 2-D tensor network. We have proposed different ways of compactification to study the effect on the roughening transition: With open boundary conditions, we find the roughening transition becomes a $\Z_2$ symmetry breaking phase transition (for odd $L_z$) and a smooth crossover (for even $L_z$). On the other hand, antiperiodic boundary conditions admit an effective $\Z_q$ clock model description with $q=2L_z$ that exhibits a BKT phase transition to a Luttinger liquid phase with an emergent $\U(1)$ symmetry and quasi long-range order. With further increasing temperature, this ordering is eventually destroyed by the bulk phase transition. From the effective-field-theory point of view, the roughening transition at $L_z\rightarrow\infty$ is essentially spontaneous symmetry breaking of emergent $U(1)$ symmetry. The continuous $U(1)$ symmetry, however, cannot be spontaneously broken due to Mermin-Wagner theorem~\cite{Mermin_1966}. The partial $\Z_2$ and $\Z_q$ symmetry breaking that we observe is thus its discrete remnant at finite $L_z$.

\par Our work is the first showcase of the power of tensor networks in this context, and the approach can be extended in many different ways. One obvious extension would consist of compressing the bond dimension of the 2-D tensor network that we need to contract, which would avoid the exponential cost with the number of layers. One of the advantages of the tensor network formalism is our ability to access the free energy directly, as compared to Monte-Carlo simulations where this typically requires a thermodynamic integration. As an illustration, in Tab.~\ref{tab:free_energy} we list some results for the interface free energy and compare with state-of-the-art Monte Carlo simulations \cite{Caselle2007}. It would be interesting to combine the free energy evaluation with a finite entanglement scaling or symmetry twist~\cite{akiyama2026} in order to obtain higher-order corrections which are relevant in several fields such as the study of confinement in gauge theories. 

\par The Ising model considered here is indeed dual to a $\Z_2$ gauge theory. The choice of antiperiodic boundary conditions corresponds to the presence of a source in the dual gauge theory, and in particular to an infinite spatial Wilson loop which can be interpreted as the leftover of a static quark anti-quark pair that has been circulated at infinity before annihilating, just leaving the corresponding Wilson line encircling the full dual lattice. The links of the original lattice orthogonal to the plane encircled by the Wilson loop are the ones responsible for the antiperiodic boundary conditions. As a result, this setup corresponds to studying the physics of Wilson loops which are relevant to understand the phases of gauge theories. In particular deep in the confined phase, the electric flux between charges should be collimated along a tube, which at low energy can be described by an effective string theory. The form of such an effective string theory has been subject of numerous studies given that it was recently shown it can be described by a $T\bar{T}$ deformation of the free boson CFT which is integrable \cite{caselle2013}. One interesting future direction is to use tensor networks and the finite entanglement scaling ansatz in order to achieve the high-precision requested to test the various sub-leading corrections to the free energy of the interface, and confirm the field theory calculations by also having a solid ground where to test higher-order finite-entanglement scaling predictions \cite{schneider2025}.

\begin{table}[]
    \centering
    \begin{tabular}{|c|c|c|}
    \hline
     & $\qquad T=2.6 \qquad$ & $T=1/0.226102$ \\
     \hline
       $\;L_z=3\;$ & $0.6188730$ & $0.0517679$ \\
       $L_z=4$ & $0.6203058$ & $0.0282305$ \\
       $L_z=5$ & $0.6204091$ & $0.0186798$ \\
       $L_z=6$ & $0.6204180$ & $0.0139667$ \\
       $L_z=7$ & $0.6204199$ & $0.0113495$ \\
       \hline
    MC & $-$ & 0.0105255(11) \cite{Caselle2007} \\
    \hline
    \end{tabular}
    \caption{The difference between the free energy density for the partition function with periodic and antiperiodic boundary conditions as a function of $L_z$ for two values of the temperature (evaluated with boundary MPS with $\chi=32$, for which these values are converged). The second value is very close to the bulk transition temperature, chosen to be able to compare with Monte Carlo data \cite{Caselle2007}.}
    \label{tab:free_energy}
\end{table}

\par In another direction, our framework also opens new avenues to study actual interfaces in 2-D quantum lattice systems \cite{Fradkin1983}, either in the context of quantum simulation experiments \cite{Balducci2023, Krinitsin2025}, lattice gauge theories in 2+1 dimensions \cite{Xu2025, DiMarcantonio2025} or interfaces in systems with (chiral) topological order. Starting from a uniform PEPS in the bulk, one would need a proper ansatz wavefunction for capturing interface states \cite{Vanderstraeten_inprep}.

\par In the context of 2D quantum models, the interface partition function as in Fig.~\ref{fig:peps} can be thought of as the euclidean version of a Loschmidt echo. For the latter, the row-to-row transfer matrix would represent a time-evolution operator of a 2-D quantum Hamiltonian. There concepts such as dynamical quantum phase transition play a central role in our understanding of the dynamics and the build-up of complexity \cite{heyl2013}. In recent works in 1D, it has indeed been shown that in absence of dynamical quantum phase transition these quantities can be calculated efficiently, because the so-called temporal entanglement obeys an area law. In presence of a dynamical quantum phase transition it still only scales logarithmically with systems size \cite{Carignano2025a, Carignano2025b}. It would be extremely interesting to explore these notions in the 2-D case as well using similar methods as the ones introduced in this work.

\parL{Acknowledgments}%
%
We would like to thank Martin Hasenbusch for an informative exchange, and Nick Bultinck and Alessio Lerose for collaborations on related projects. L.T. would like to thank M. Caselle  F. Gliozzi, and M. Panero for pointing out several relevant references on the subject.  A.U. is supported by FWO Junior Postdoctoral Fellowship (grant No. 3E0.2025.0049.01) and Watanabe foundation. L.B. is supported by European Research Council (ERC) under the European Union’s Horizon 2020 program (GaMaTeN, grant agreement No. 101125822). L.T. is supported by the "Proyecto PID2024-160172NB-I00 founded by the MICIU/AEI/10.13039/501100011033 and the EU scheme FEDER.

\bibliography{bibliography.bib}

\end{document}